# Lesion Net - Skin Lesion Segmentation Using Coordinate Convolution and Deep Residual Units

Sabari Nathan, Priya Kansal
Couger Inc., Japan
sabari@couger.co.jp, priya@couger.co.jp

**Abstract:** Skin lesions segmentation is an important step in the process of automated diagnosis of the skin melanoma. However, the accuracy of segmenting melanomas skin lesions is quite a challenging task due to less data for training, irregular shapes, unclear boundaries, and different skin colors. Our proposed approach helps in improving the accuracy of skin lesion segmentation. Firstly, we have introduced the coordinate convolutional layer before passing the input image into the encoder. This layer helps the network to decide on the features related to translation invariance which further improves the generalization capacity of the model. Secondly, we have leveraged the properties of deep residual units along with the convolutional layers. At last, instead of using only cross-entropy or Dice-loss, we have combined the two-loss functions to optimize the training metrics which helps in converging the loss more quickly and smoothly. After training and validating the proposed model on ISIC 2018 (60% as train set + 20% as validation set), we tested the robustness of our trained model on various other datasets like ISIC 2018 (20% as test-set) ISIC 2017, 2016 and PH2 dataset. The results show that the proposed model either outperform or at par with the existing skin lesion segmentation methods.

## 1. Introduction

The incidences of malignant melanoma are rapidly increasing worldwide in recent years [1][2][3]. The skin lesions are the only symptoms of melanoma. The most used method to identify Melanomas skin lesions is dermatology imaging method which is followed by human vision and observation. However, the involvement of human vision and observation sometimes is not that accurate and becomes subjective also [8][19]. Thus, there is a requirement of some automated and unbiased system which can identify the dangerous skin lesions that cause melanoma. However, the automated skin lesion segmentation is not an easy task because of many reasons which include the availability of data for training, irregular shapes, unclear boundaries, and different skin colors. Sometimes, the images have very low contrast because of which it is difficult to differentiate between lesion and skin. Also, in some images, the lesion part is covered with the hairs, frame, blood vessels, air bubbles, etc., which make the lesion segmentation the task more challenging. Some sample images from ISIC 2018 and ISIC 2016 datasets are shown in Figure 1.

In the previous works, the researchers have majorly discussed the use of thresholding-based method, Region-Based Methods, Fusion Based Methods and Deformable Models for the medical image segmentation. However, the impressive success of the deep learning-based approaches in semantic segmentation has attracted the researchers in the field of medical image segmentation also. Our proposed framework is also a deep learning-based approach which follows an encoding-decoding structure without any prior knowledge of input images. In the encoding phase, we have used deep residual blocks to extract information about lesion skin or normal skin. Before passing the input into the residual blocks, we have fed the input in a coordinate convolution layer. The same residual blocks are also used in the decoder part. Finally, the weights are optimized with the help of an average of dice loss and cross entropy loss. Our main contributions in this work are:

1. Application of coordinate convolutional layer before the encoder
2. Deep Residual units to extract the more detailed information about the semantic segmentation which helps the model to learn more about the boundary information.
3. Customized loss function which is an average of the dice loss function and cross entropy loss function which has helped quick and smooth optimization.

Our model has the ability to learn rich hierarchical and contextual features. The extensive experiments on the four public datasets, ISIC 2018, ISIC 2107, ISIC 2016 and PH2 datasets prove that our model in a robust model which outperform or equally perform the state of art on all the public datasets. The rest of the paper discusses the literature, the details of the proposed architecture experiments, results and conclusion.

## 2. Literature Review

A considerable amount of research work on developing an automated system for lesion segmentation is available. The researchers have developed a range of segmentation techniques. A very popular technique is thresholding-based method [4][5] in which the lesion area is segmented by comparing with a threshold value. These techniques perform well in low-level segmentation task. But sometimes, region-based segmentation leads to over-segmented areas especially in case of low contrast, diverse color and unclear boundaries in the image. Another important approach for segmentation is Region Based Methods [6][7][8]. The regions-based approaches are based

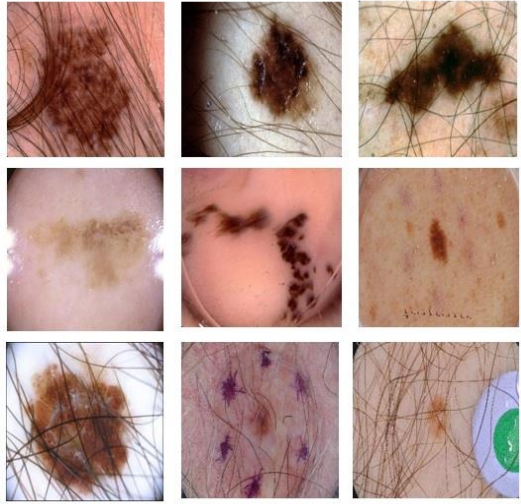

FIGURE1: Sample Images from ISIC 2018 and 2016 datasets

on the fact that the nearby pixels have homogeneous color values. Later, researchers have also experimented some fusion based approaches [9][10]. These fusion techniques are comparatively easier and provide good results except when there is a high variability in centre and boundaries of the lesion region [11]. A review of the state of art of these methods is discussed in [20]. However, with the transformations in AI and the success of deep-learning based methods in image processing task, now researchers have started using deep neural networks for the purpose of medical image segmentation. For instance, the classical convolutional neural networks (e.g. Inception, VGG Net, and ResNet) have been widely applied for the impressive performance [12][15] in the skin lesion segmentation. [16][17][14] have even used the pre-trained weights (trained on ImageNet Dataset) of these models extracting the features of dermoscopic images. Similarly, [18][24][35] have used different models and ensemble the results using bagging strategy to improve the accuracy of segmentation. Ensemble strategies have been very striking in the past few years. [14][21][22] have ensemble of different architectures used for extracting data features, which are further inputted to the SVM classifier (deep learning combined with SVM is also used in other approaches [23]). However, since the inception of U-Net architecture [19][20], it has been proved to be a backbone structure for the state of art results in segmentation tasks. The winners of ISIC 2018 [35] challenge have used the U-net structure with Resnet 101 as the backbone of the encoder to extract the features.

In the present work, we have also used the U Net structure. As a backbone of this encoder-decoder network, we have used the deep residual units. We have also used coordinate convolution layer before feeding the image into the encoder, hence the encoding has started with the 5 channel output of coordinate convolutional layer instead of 3 channel original input image.

## 3. Lesion Net

In our proposed Lesion Net, the skin lesion segmentation is carried out with the help of U-Net network [19][20]. The U-Net has two modules: Encoder and Decoder.

Figure 2 shows the detailed architecture. The encoder module is used for down sampling and feature extraction from the input image. The decoder is used to up sample the image features and creating the segmentation map. We have used four down sampling and four up sampling blocks. As a backbone of encoder and decoder, we used deep residual units inspired by [25]. Every down sampling sampling block consists of one convolutional layer and two deep residual units. The output of each down sampling block is fed into the following block and also concatenated with the corresponding up sampling block using skip connections. In the decoder, each up-sampling block consist of the deconvolution layer concatenated with the output of the corresponding down sampling block followed by two residual units. Lastly, the output of the decoder is sent to the last convolutional layer with SoftMax activation function, which works as a classifier and generates the probability for each pixel for background and foreground independently. Also, before this passing the input image to the encoder, the image is sent to the coordinate convolutional layer. Further details are given in the following section.

### 3.1. Coordinate Convolutional Layer

As shown in Figure 3, before passing the image into the encoder, we have first passed the input image into a coordinate convolution layer as proposed by [26]. Coordinate convolutional layer helps the network to decide on the features related to translation invariance which further improves the generalization capacity of the model. As suggested in the original paper, with the help of coordinate convolutional layer, spatial coordinates can be mapped with the coordinates in Cartesian space through the use of extra coordinate channels which gives the power to the model to use either complete or varying degree of translation features which is quite helpful in improving the translation invariance and thus the generalization capacity. Here, we have used the two extra coordinate channels similar to the original paper. The first coordinate channel is a matrix in which row one is filled with all zeros, row 2 is all 1s, row three is all 2s and so on wherein the other channel the numbers are filled in the column.

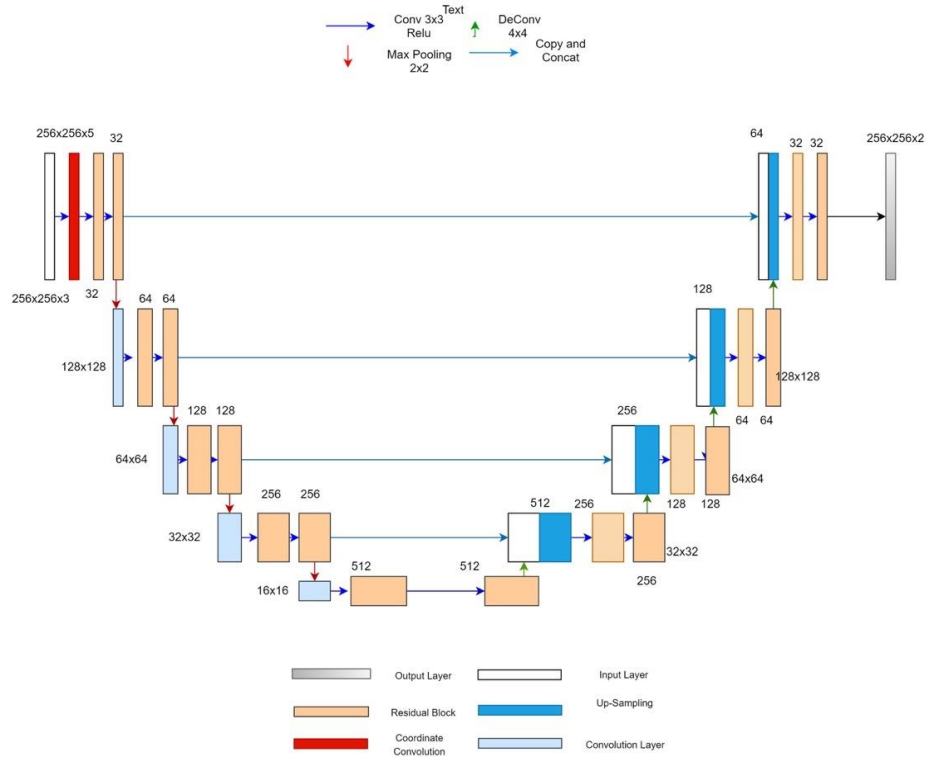

FIGURE 2: Proposed Architecture: Lesion-Net which includes the encoder-decoder structure of U-Net

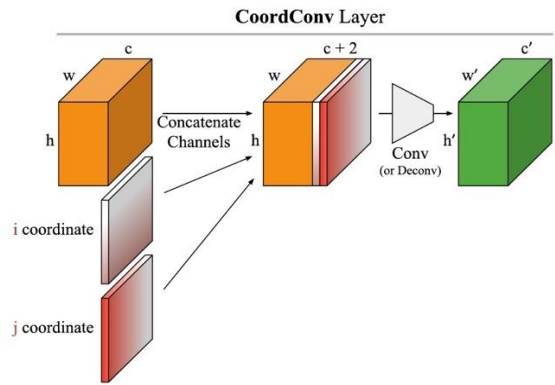

FIGURE 3: A typical residual block used in Lesion Net

### 3.2. Deep Residual Unit

The number of layers in the neural network plays an important role in the performance of the model. In fact, this has been a proven fact now that the deeper the network, the higher the performance, because the pile of layers helps the network to extract the hierarchical features more efficiently [27][28]. However, if we simply stack the layers, it can also lead to the problem of vanishing gradient and lead to a non-converging training process or converging at the local optimal minima instead of global optimal minima. There is also a likelihood that increasing the number of layers will lead to a sharp convergence, which reduces the generalization capacity of the network.

A deep residual unit as inspired by [25] is an effective framework in resolving the problem of vanishing gradient and sharp convergence as it maps the input with the output of following convolutional layers with the help of skip connections. This eases the process of backpropagating as it makes the gradient transfer back to the initial layers or to the preceding block. The residual units also have the leverage of reduced computational complexity of the model to some extent as use of the skip connections keeps the parameters unchanged.

Our network has deep residual units as the backbone of encoder and decoder. Similar to [25], we have also used three convolutional layers in the deep residual units. However, as shown in Figure 4, in place of changing the number of filters and strides, we have kept them constant in the entire unit. Similar to the original residual unit proposed in Resnet 34, we have used same 3X3 strides for each convolutional layer in all residual units throughout the network and same filter count (starting from 32 to 512 in encoder and from 512 to 32 in the decoder) in a particular up sampling or down sampling residual unit throughout the network. Each convolutional layer is followed by batch normalization and activation to deal with the problem of vanishing gradient and overfitting. Figure 3 shows the detail of the residual unit used in the architecture.

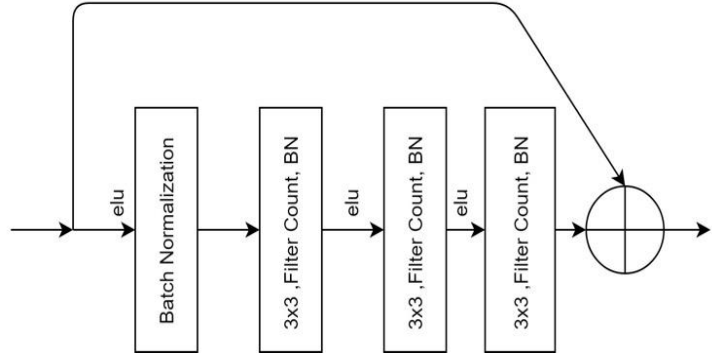

FIGURE 4: A typical residual block used in Lesion Net

The output of residual blocks is fed to the next block and to the corresponding block in the decoder through skip connection. The reason to use the multiple inputs in up sampling blocks of the decoder module is that it can restore the multiple resolution information as features. The restored multiple features can be fused to enhance segmentation performance.

## 4. Experimental Setup and Results

### 4.1. Training Data

ISIC 2018 [10][11] is a subset of the large International Skin Imaging Collaboration (ISIC) archive, which contains dermoscopic images acquired on a variety of different devices at numerous leading international clinical centers. The ISIC 2018 skin lesion segmentation challenge dataset provides 2,594 training images. Image size varied from 540X576 pixels to 4499X6748 pixels. Clinical experts provided manual delineations in the training data. Also, for evaluation and testing purpose, we further randomly split the training images into 60-20-20 ratio. Thus, the training, validation and test images set consists of 1556, 520, 518 respectively. To test the robustness and generalization of our trained model, we have also tested ISIC 2016, ISIC 2017 and PH2 dataset using the trained model with proposed architecture and loss function. The details of the other datasets are as follows:

*4.1.1 ISIC 2016 Test Dataset* This is the test data released by International Skin Imaging Collaboration (ISIC) in 2016 which contains 379 images. The images contain a representative mix of images of both malignant and benign skin lesion [12].

*4.1.2. ISIC 2017 Test Dataset* This dataset also contains the 600 bio-medical images issued for Lesion Image segmentation by ISIC (International Skin Imaging Collaboration) in 2017 [10].

*4.1.3 PH2 Dataset* This dataset includes 200 dermoscopic images that were obtained at the Dermatology Service of Hospital Pedro Hispano (Matosinhos, Portugal) [25].

### 4.2. Data Pre-processing and Augmentation

Although our proposed model does not require any special preprocessing or post processing, but to speed up the training process, we downsized the images. Unlike the original U Net, we provide an input of 256 X 256 pixels with 3 channels and get an output in the form of mask (lesion region), the boundary of the lesion and background image. The Lesion-Net loss is converged based on proposed loss. An additional set of transformations were applied to the images at runtime. These transformations included the horizontal flipping, vertical flipping, rotation, shearing and zoom.

### 4.3. Evaluation Metrics

In this paper, we evaluate the segmentation performance by following the assessment criteria of the ISIC 2016, 2017 and 2018 Lesion Segmentation Challenge [33][34][36], which includes Dice coefficient (DC), Jaccard Index (JA), Accuracy (AC), Sensitivity (SE), and Specificity (SP). JA and DC are similarity metrics that measure the overlapping between the predicted results and the ground truth.

### 4.4. Loss Functions

As stated earlier, we have used an average of two loss function viz. Dice loss and cross entropy loss. This average loss is used to update the weights while back propagation process with the Adam.

$$Loss = L * 0.5 + Dice\ Loss * 0.5$$

$$Dice\ Loss = 1 - \frac{2 * \sum_{n=k}^{i=0} y_i p_i + \epsilon}{\sum_{n=k}^{i=0} y_i + \sum_{n=k}^{i=0} p_i + \epsilon}$$

$$L = -\sum_{j=0}^{M} \sum_{i=0}^{N} y_{ij} \log(p_{ij})$$

TABLE 1: Segmentation results on the different dataset with Coordinate Convolutional Layer (W-C) and without Coordinate Convolutional Layer (WO-C)

| Dataset | Dice | | Jaccard | | Sensitivity | | Specificity | |
|---|---|---|---|---|---|---|---|---|
| | W-C | WOC | W-C | WOC | W-C | WOC | W-C | WOC |
| ISIC 2018 Train | 0.92 | 0.915 | 0.8602 | 0.8529 | 0.9297 | 0.9242 | 0.9742 | 0.9799 |
| ISIC 2018 Val | 0.8423 | 0.8278 | 0.7612 | 0.7431 | 0.8827 | 0.8696 | 0.9758 | 0.9823 |
| ISIC 2018 Test | 0.8997 | 0.8825 | 0.8325 | 0.8119 | 0.9185 | 0.8829 | 0.9456 | 0.9713 |
| ISIC 2017 Test | 0.8787 | 0.8392 | 0.7828 | 0.753 | 0.8623 | 0.8302 | 0.9608 | 0.9785 |
| ISIC 2016 Test | 0.9239 | 0.9166 | 0.8647 | 0.855 | 0.9362 | 0.9267 | 0.9645 | 0.9703 |
| PH2 Dataset | 0.9087 | 0.8832 | 0.8384 | 0.8068 | 0.9763 | 0.9241 | 0.9188 | 0.946 |

### 4.5. Training

There is no pre-trained weight used in the training of the proposed model. The network is trained from scratch using the Adam optimization [32] algorithm. As stated earlier also, the loss function is the average loss function. The learning rate is initialized with 0.001 and reduced after 10 epochs to 10% if validation loss does not improve. The model is trained for 100 epochs with a mini-batch size is 4 and validation are done after every 10 steps. We used the validation loss in call back function to save the best model weights. The dataset is trained using Nvidia 1080 GTX GPU.

### 4.6. Results

*4.6.1 Experiment on the coordinate Convolutional Layers* Since we have used the coordinate convolution layer as the very first layer of the proposed model to improve the generalization capacity of the model. For the purpose of demonstrating the effectiveness of Coordinate convolution layer, we have trained the proposed model, with and without coordinate convolutional layer. Table 1 shows that the performance of the model when using the coordinate convolutional layer is significantly higher than the case when we have not implemented the coordinate convolutional layer. W-CCL means that the results are with the coordinate convolutional layers and WO-CCL means that the results are without the coordinate convolutional layer. Notably, the performance of the model with coordinate convolutional layer is high for all the datasets in terms of all the metrics. The reason for that is more generalization capacity of the model which helped the model, which is trained only on ISIC 2018, to perform on all the datasets.

TABLE 2: Segmentation results on the different dataset with three layers (3-L) and two layers (2-L)

| Dataset | Dice | | Jaccard | | Sensitivity | | Specificity | |
|---|---|---|---|---|---|---|---|---|
| | 3-L | 2L | 3-L | 2L | 3-L | 2L | 3-L | 2L |
| ISIC 2018 Train | 0.92 | 0.915 | 0.8602 | 0.8529 | 0.9297 | 0.9242 | 0.9742 | 0.9799 |
| ISIC 2018 Val | 0.8423 | 0.8278 | 0.7612 | 0.7431 | 0.8827 | 0.8696 | 0.9758 | 0.9823 |
| ISIC 2018 Test | 0.8997 | 0.8981 | 0.8325 | 0.8206 | 0.9185 | 0.9013 | 0.9456 | 0.962 |
| ISIC 2017 Test | 0.8787 | 0.8491 | 0.7828 | 0.7645 | 0.8623 | 0.8422 | 0.9608 | 0.9719 |
| ISIC 2016 Test | 0.9239 | 0.9167 | 0.8647 | 0.8543 | 0.9362 | 0.9177 | 0.9645 | 0.9741 |
| PH2 Dataset | 0.9087 | 0.9149 | 0.8384 | 0.8495 | 0.9763 | 0.9585 | 0.9188 | 0.9413 |

*4.6.2 Experiment on the layers in Residual units* We have also conducted an experiment to evaluate the impact of layers in residual units. We have conducted the experiments to compare the 2 layers as suggested in the original block and three layers as we used. **Table 2** shows the model results with 3 layers and 2 layers in the residual units. It is obvious from the table that use of three layers in residual units is helping the model to extract deeper hierarchical and contextual features and thus to be more accurate.

*4.6.3 Experiment on the Loss Function* We have performed experiments with the loss function also. We have trained the proposed architecture with the proposed custom loss function as well as with the cross-entropy loss. **Table 3** demonstrates the various evaluation metrics for the loss function. The proposed custom loss has not only helped the model in converging smoothly but also scored better number when it comes to evaluation metrics. The average of cross entropy and the Dice loss helps in alleviating the backpropagation process and try to help the model not to deviate from the way of global optimal minima.

*4.6.4 Comparison with other Methods* In order to show the effectiveness and robustness of our proposed model, we have compared our results with some state of art methods for all the datasets. **Table 4** shows that the proposed method with the coordinate convolutional layer and the deep residual unit has achieved superior performance than other methods for all the datasets. We have mentioned the results of the approaches used by the winners of the ISIC 2016, 2017, 2018 challenge along with some other state of art results. In, 2018 the winner has used the same U-Net kind of structure with RESNET as the backbone and some multiscale blocks before up sampling. The winner of 2017 [42] has used some dual threshold-based method to extract the features and then RESNET in up sampling block to generate the segmentation mask. However, our proposed method is not only better but robust also compare to other methods.

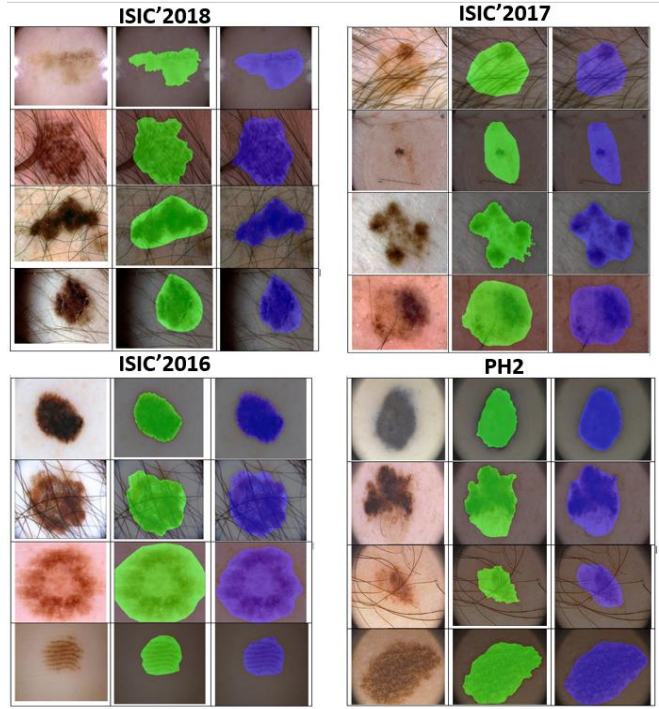

FIGURE 5: Results of Lesion Net for four Datasets: the first column of each block (dataset) shows the input image, second column shows the ground-truth and the third column shows the segmentation results of Lesion Net

TABLE 3: Segmentation results on the different dataset with Custom Loss (CL) and Cross Entropy (CE)

| Dataset | Dice | | Jaccard | | Sensitivity | | Specificity | |
|---|---|---|---|---|---|---|---|---|
| | CL | CE | CL | CE | CL | CE | CL | CE |
| ISIC 2018 Train | 0.9200 | 0.9025 | 0.8602 | 0.8356 | 0.9297 | 0.9190 | 0.9742 | 0.9766 |
| ISIC 2018 Val | 0.8423 | 0.8324 | 0.7612 | 0.7516 | 0.8827 | 0.8789 | 0.9758 | 0.9796 |
| ISIC 2018 Test | 0.8997 | 0.8962 | 0.8325 | 0.8280 | 0.9185 | 0.9237 | 0.9456 | 0.9466 |
| ISIC 2017 Test | 0.8787 | 0.8419 | 0.7828 | 0.7557 | 0.8623 | 0.8612 | 0.9608 | 0.9631 |
| ISIC 2016 Test | 0.9239 | 0.9072 | 0.8647 | 0.8407 | 0.9362 | 0.9182 | 0.9645 | 0.9684 |
| PH2 Dataset | 0.9087 | 0.9100 | 0.8384 | 0.8420 | 0.9763 | 0.9729 | 0.9188 | 0.9182 |

TABLE 4: Performance Comparison on Various datasets among Lesion Net and Other models.

| Dataset | Algorithm | Dice | JAC |
|---|---|---|---|
| ISBI 2016 | ExB [33] | 0.91 | 0.843 |
| | CUMED [33] | 0.897 | 0.829 |
| | Mahmudur [33] | 0.895 | 0.822 |
| | **Lesion Net (Ours)** | **0.9239** | **0.8647** |
| ISBI 2017 | Yuan et al.[42] | 0.849 | 0.765 |
| | Berseth et al. [39] | 0.847 | 0.762 |
| | Bi et. al. [38] | 0.844 | 0.76 |
| | **Lesion Net (Ours)** | **0.8787** | **0.7828** |
| ISBI 2018 | Chengyao Qian[36] | 0.898 | 0.838 |
| | Youngseok0001[36] | 0.904 | 0.837 |
| | **Lesion Net (Ours)** | **0.8997** | **0.8325** |
| PH2 | Ahn et. al. [37] | 0.6997 | 0.577 |
| | **Lesion Net (Ours)** | **0.9087** | **0.8384** |

## 5. Discussion and Conclusion

In order to enhance the accuracy and robustness, we have proposed an architecture which leverage the benefits of coordinate convolutional layer and deep residual block in an encoding-decoding framework for skin lesion segmentation. To demonstrate the results of our proposed model, we have displayed the resulted segmentation of some sample images from 2017 and 2018 datasets in Figure 5 and compare those with the ground truth images. Figure 5 shows that our model is good enough not only to create a clear segmentation mask but to restore the original image. We have evaluated our results on the basis of the following facts:
1. We have not used any pre-trained weights to extract the features.
2. Less computationally expensive because of the use of residual block
3. More robust because of the use of coordinate convolutional layer
4. More smooth and quick training process because of the customized loss function

However, in the future, we will try replacing the residual units with some other custom units and also will try to combine some handcrafted features or threshold-based features along with the CNN layers features.